\documentclass[reprint, cha]{revtex4-1}

\usepackage{amssymb}
\usepackage[english]{babel}    % para texto em Inglês
\usepackage[dvips]{graphics}
\usepackage{subfigure}
\usepackage[pdftex]{graphicx}
\usepackage{multirow}
\usepackage{rotating}
\usepackage{picinpar}
\usepackage{color}
\usepackage{lineno}
\usepackage{listings}

\begin{document}

\title{Characterization of nanostructured material images using fractal descriptors}

\author{Jo\~{a}o B. Florindo}
 	     \email{jbflorindo@ursa.ifsc.usp.br}
\affiliation{Instituto de F\'{i}sica de S\~{a}o Carlos - Universidade de S\~{a}o Paulo - S\~{a}o Carlos - SP - Brazil} 

\author{Mariana S. Sikora}
 	     \email{ma\_sikora@hotmail.com}
\affiliation{Interdisciplinary Electrochemistry and Ceramics Laboratory, Federal University of S\~{a}o Carlos, S\~{a}o Carlos, Brazil} 

\author{Ernesto C. Pereira}
 	     \email{decp@power.ufscar.br}
\affiliation{Interdisciplinary Electrochemistry and Ceramics Laboratory, Federal University of S\~{a}o Carlos, S\~{a}o Carlos, Brazil} 

\author{Odemir M. Bruno}
              \email{bruno@ifsc.usp.br}
\affiliation{Instituto de F\'{i}sica de S\~{a}o Carlos - Universidade de S\~{a}o Paulo - S\~{a}o Carlos - SP - Brazil}

\date{\today}

\begin{abstract}
This work presents a methodology to the morphology analysis and characterization of nanostructured material images acquired from FEG-SEM (Field Emission Gun-Scanning Electron Microscopy) technique. The metrics were extracted from the image texture (mathematical surface) by the volumetric fractal descriptors, a methodology based on the Bouligand-Minkowski fractal dimension, which considers the properties of the Minkowski dilation of the surface points.
An experiment with galvanostatic anodic titanium oxide samples prepared in oxalyc acid solution using different conditions of applied current, oxalyc acid concentration and solution temperature was performed. The results demonstrate that the approach is capable of characterizing complex morphology characteristics such as those present in the anodic titanium oxide. 
\end{abstract}


\maketitle

\section{Introduction}

The morphology analysis of solid samples is an important research area in Materials Science to characterize some of their properties \citep{MF95,MOANNT99,YNNFM02,KGAS08}. Generally, the technique employed in this kind of application consists in the  quantitative analysis of the micrographs obtained using one (or more) of following techniques, such as FEG-SEM (Field Emission Gun-Scanning Electron Microscopy), AFM (Atomic Force Microscopy), STM (Scanning Tunneling Microscopy) or TEM (Transmission Electron Microscopy). In any case, the results are a matrix of values which expresses the topography of the measured sample. 

When the morphology of the material is investigated, we observe that each image presents a specific distribution pattern. This distribution is quite similar to that found in textures classically studied with image analysis tools, e.g., the so-called texture analysis methods \citep{K99}. Therefore, the application of this procedure over, for example, FEG-SEM images, could make possible the modeling of physical important properties of the materials, such as roughness, grain boundaries, morphologic defects and total surface area. The quantitative analysis of nanostructured materials for well behaved samples has been described using simple functions of existing software packages. One example is the automatic counting of pores in self-organized anodic porous alumina \citep{MF95} using ImageJ \citep{AMR04} and Gwyddion \citep{NK12}. The problem is different when studying complex  morphology characteristics of the samples as those ones where the distinction among the patterns is not obvious. In this case, a more sophisticated analysis must be used and these ones generally are not included in those software packages described above. 

Among the methods described in the literature, those ones based on fractal analysis have presented excellent performance in the investigation of complex textures, mainly on those synthesized and natural samples \citep{BCB09, BPFC08, FCB10, PPFVOB05}. Actually, nature is rich in self-similar patterns, that is, structures which repeat themselves under different scales. From a mathematical point of view, this is also an intrinsic property of fractal objects. Therefore, fractal geometry is appropriate to measure such kind of structures and, as consequence, self-similarity also measures complexity (meaning the level of details along scales) which is directly related to spatial occupation in the structure. 

Notwithstand that fractal dimension provides a good solution in many object identification problems, it is limited in the representation of some classes of objects \cite{FB11} due to two main aspects: i) fractal dimension is a real number which is insufficient to characterize  such objects. ii) there are samples that have different patterns but present the same fractal dimension. In fact, these objects can have a more or less self-similar aspect depending on the scale observed \citep{M68}. In the literature, different propositions have been presented to solve the above drawbacks. The main ones are multifractal \citep{H01}, multiscale fractal dimension \citep{MCSM02} and fractal descriptors \citep{BPFC08}. Considering that several papers have demonstrated the superior perfomance of fractal descriptors dealing with texture images over the other ones, we are focused here on such approach as a tool for texture discrimination \citep{BPFC08,FCB10,K99,PPFVOB05}. 

We apply Volumetric Minkowski descriptors methodology, which was initially developed in \citep{BCB09}. It is derived from the Bouligand-Minkowski fractal dimension. The descriptors are obtained by mapping the original gray level image (of FEG-SEM data, in this case) onto a three-dimensional mathematical surface. Thus, such surface is  dilated by Bouligand-Minkowski method using spheres with predefined radii. The fractal descriptors are then estimated from the volume of dilation for each sphere radius. With the growing of the dilation radius, the spheres start to interfere among themselves, forming a wavefront which is tightly related to the structure of the material. It is important to stress out that the dilation process captures the arrangement of the topography \cite{BCB09}. Thus, these descriptors are capable of providing very rich information about the morphology of the material and, consequently, are a strong method for nanostructured material characterization task. The use of the Volumetric Fractal Descriptors applied to nano structured surfaces was initially proposed in \cite{FSPB12}. In
that seminal project, was suggested that Fractal Volumetric Descriptors could be used to characterize and analyze such nanostructures, showing the discrimination power on two distinct conditions. In the present work, an experiment with galvanostatic anodic titanium oxide samples prepared in oxalyc acid solution using eight different conditions of applied current, oxalyc acid concentration and solution temperature was performed, and demonstrate that the proposed technique is capable to identify the nanosurfaces. The nature of the materials surfaces and its images are a difficult problem of image analysis, and the proposed technique demonstrate be suitable to characterize and to identify nanostructured surfaces.

%Considering the exposed above, in this work we develop an analysis method to characterize the morphology of anodic titanium oxide samples using a Volumetric Minkowski descriptors approach. 

\section{Materials and Methods}

\subsection{Materials} \label{sec:mat}

\begin{figure*}
\centering
	\begin{tabular}{cccc}
		\includegraphics[width=0.5\textwidth]{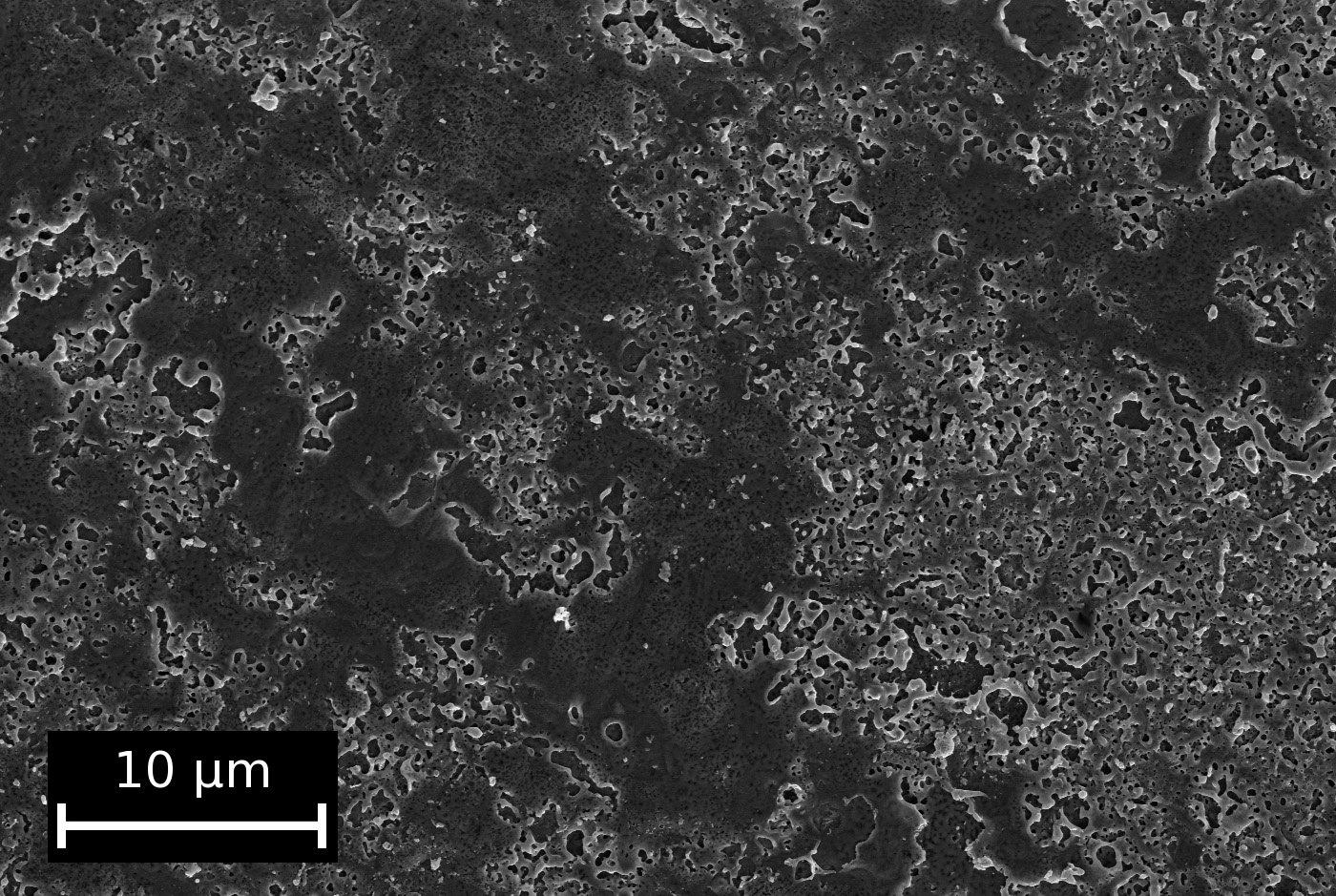}(1) & \includegraphics[width=0.5\textwidth]{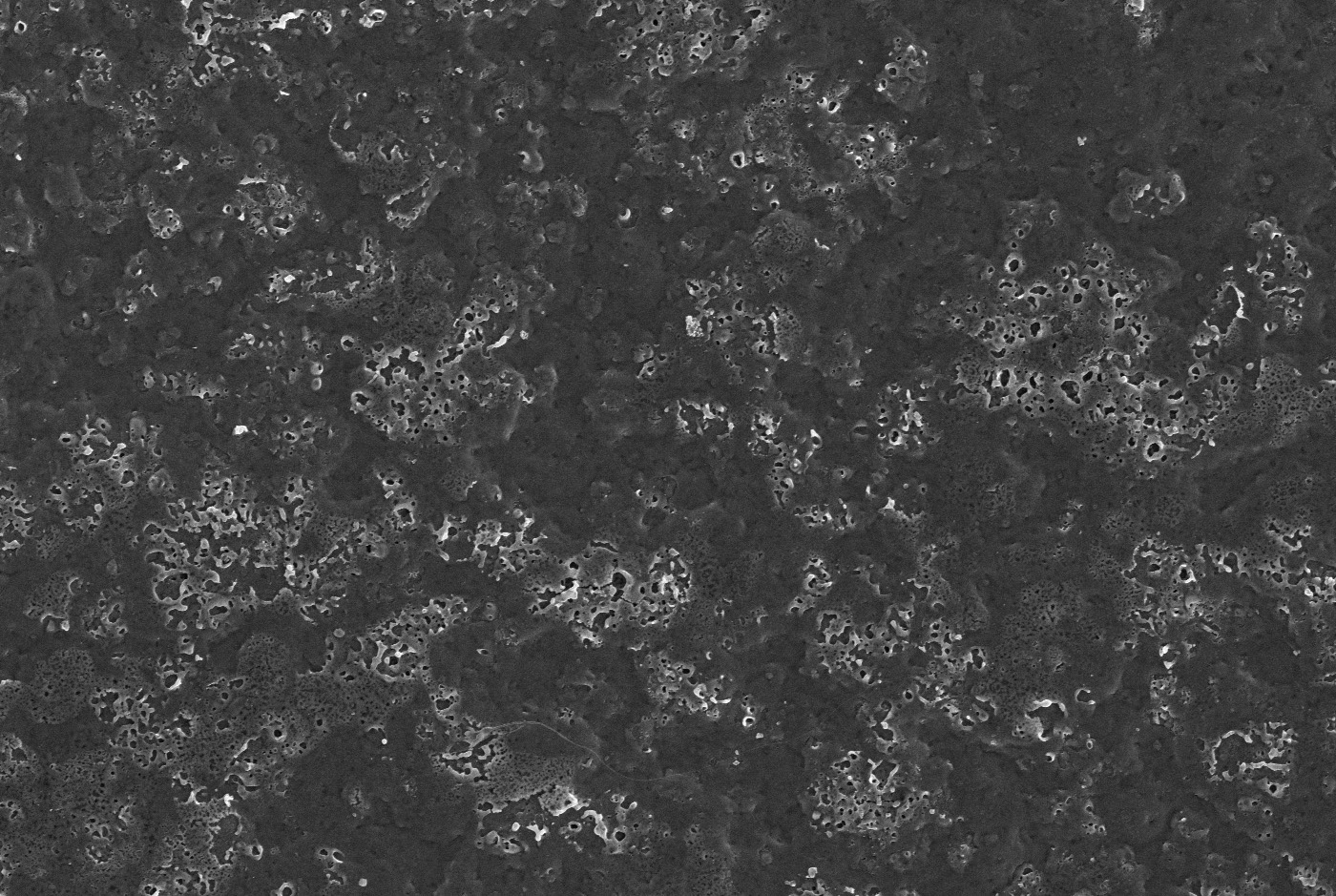}(2)\\
		 \includegraphics[width=0.5\textwidth]{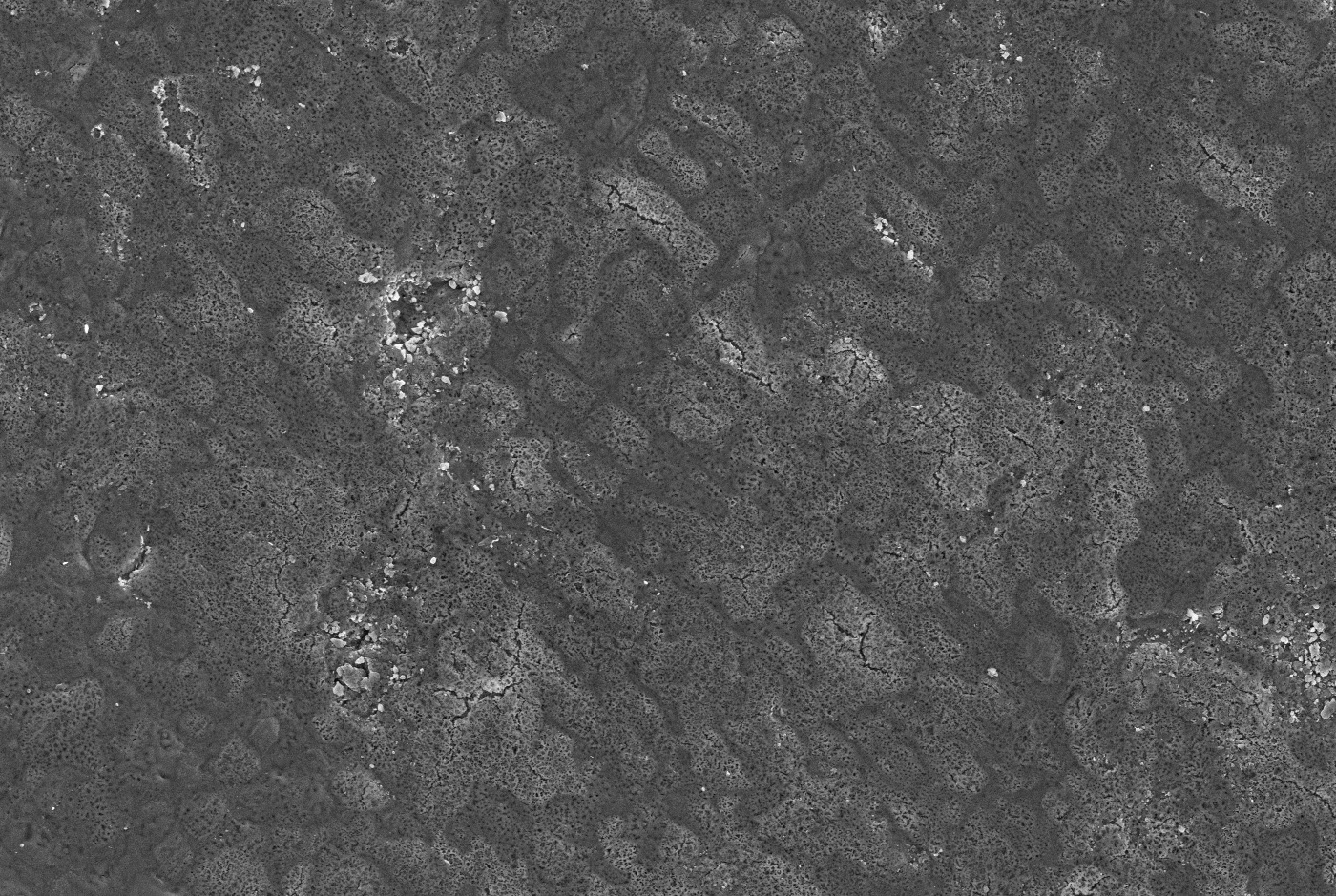}(3) & \includegraphics[width=0.5\textwidth]{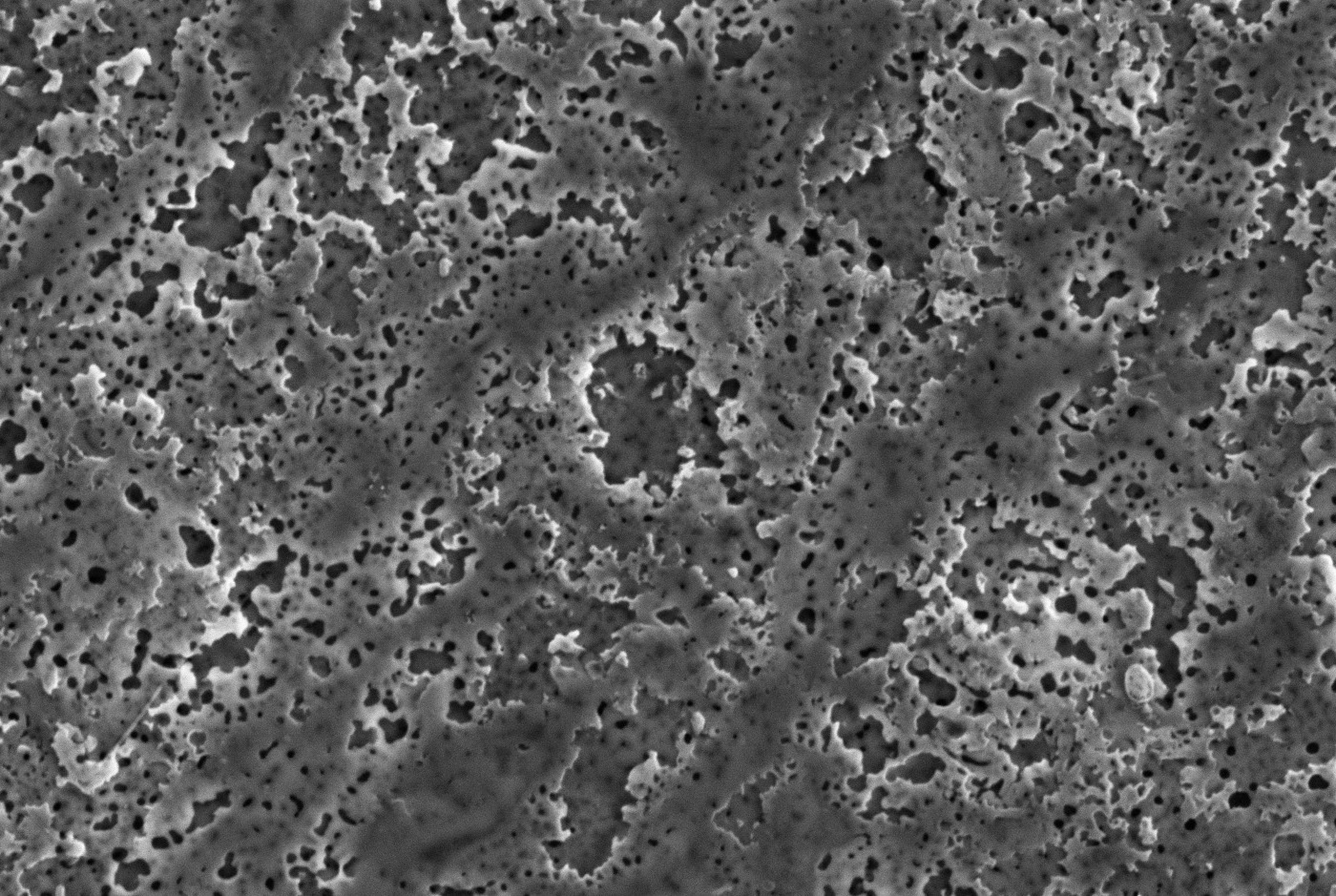}(4)\\
		\includegraphics[width=0.5\textwidth]{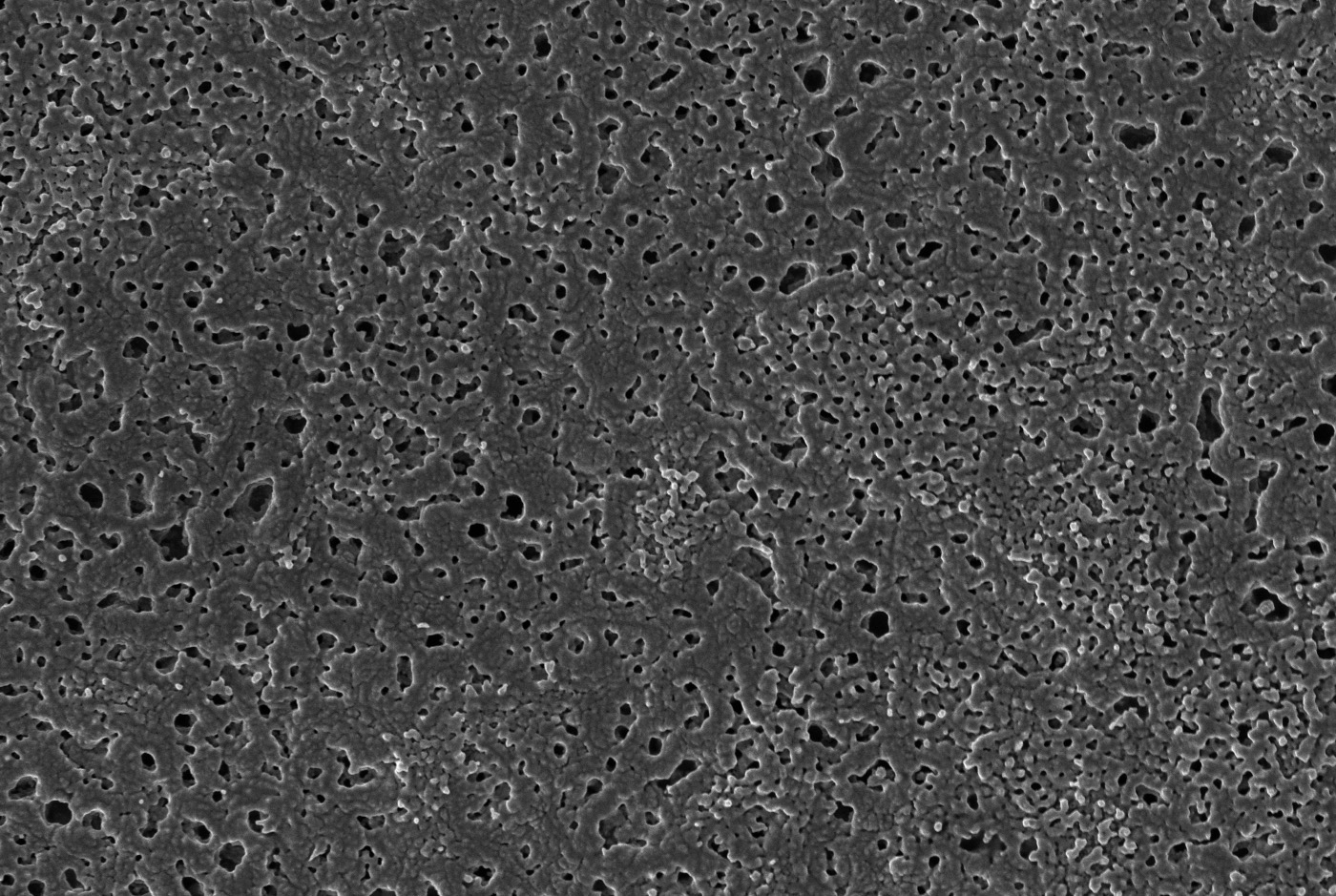}(5) & \includegraphics[width=0.5\textwidth]{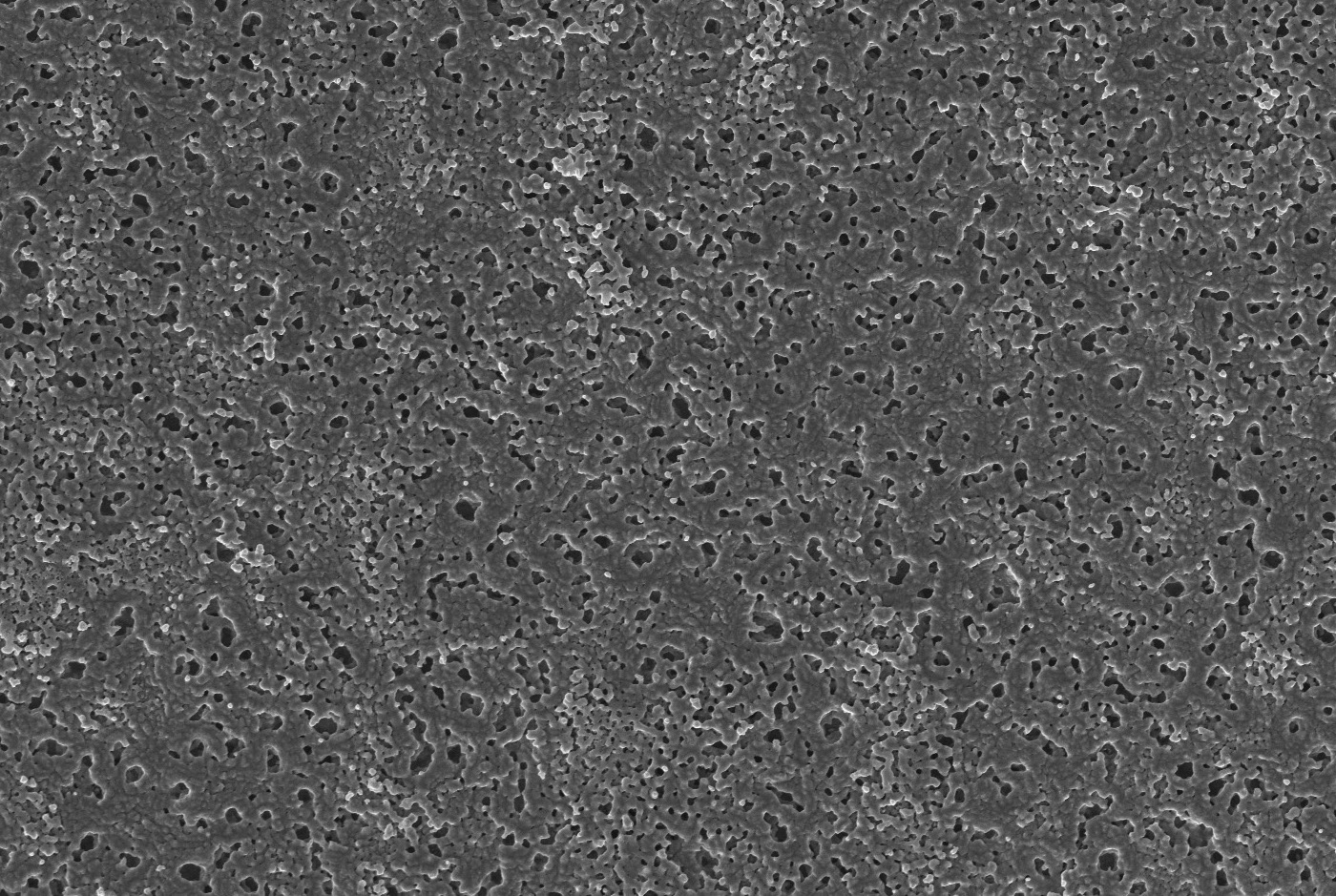}(6)\\
		 \includegraphics[width=0.5\textwidth]{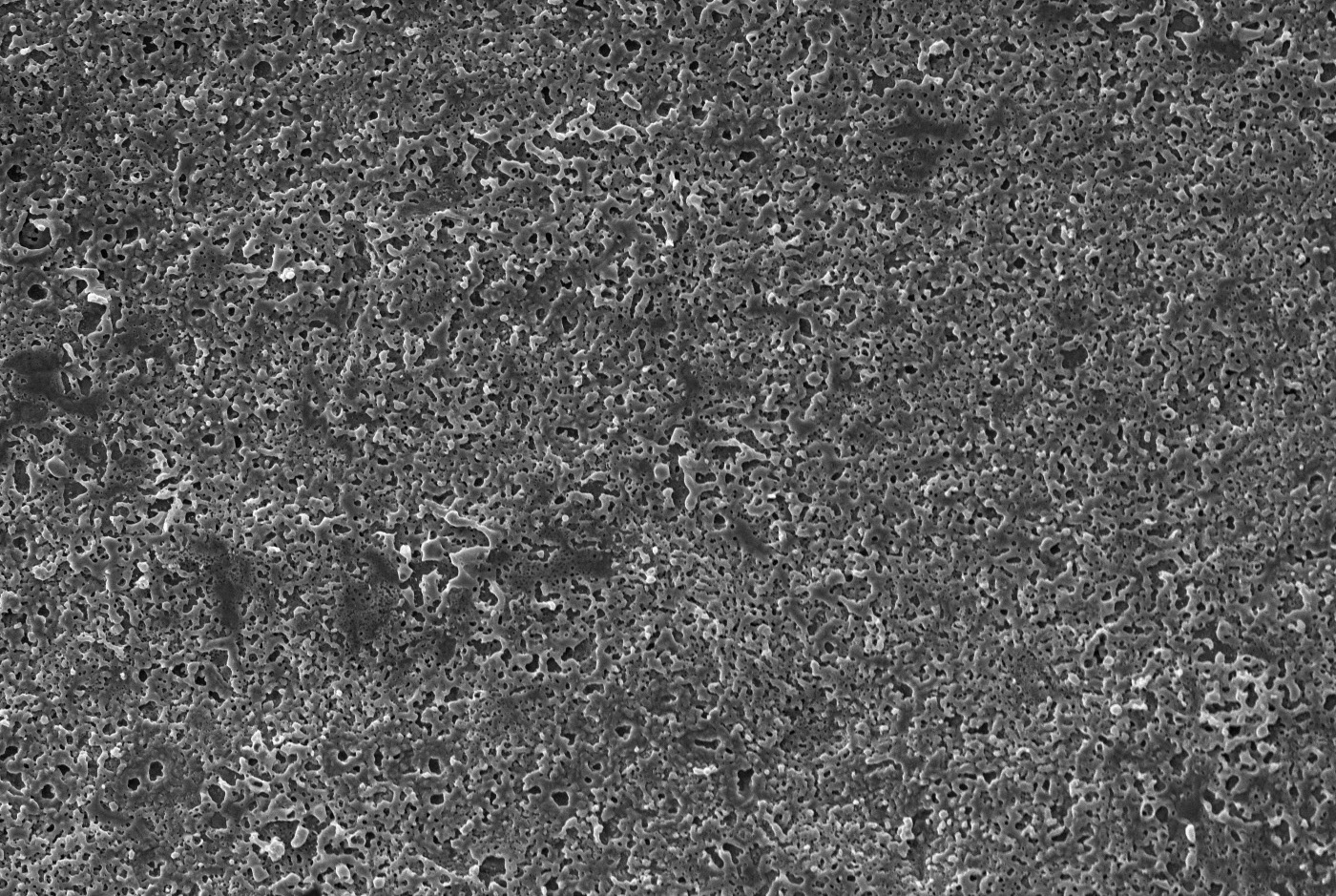}(7) & \includegraphics[width=0.5\textwidth]{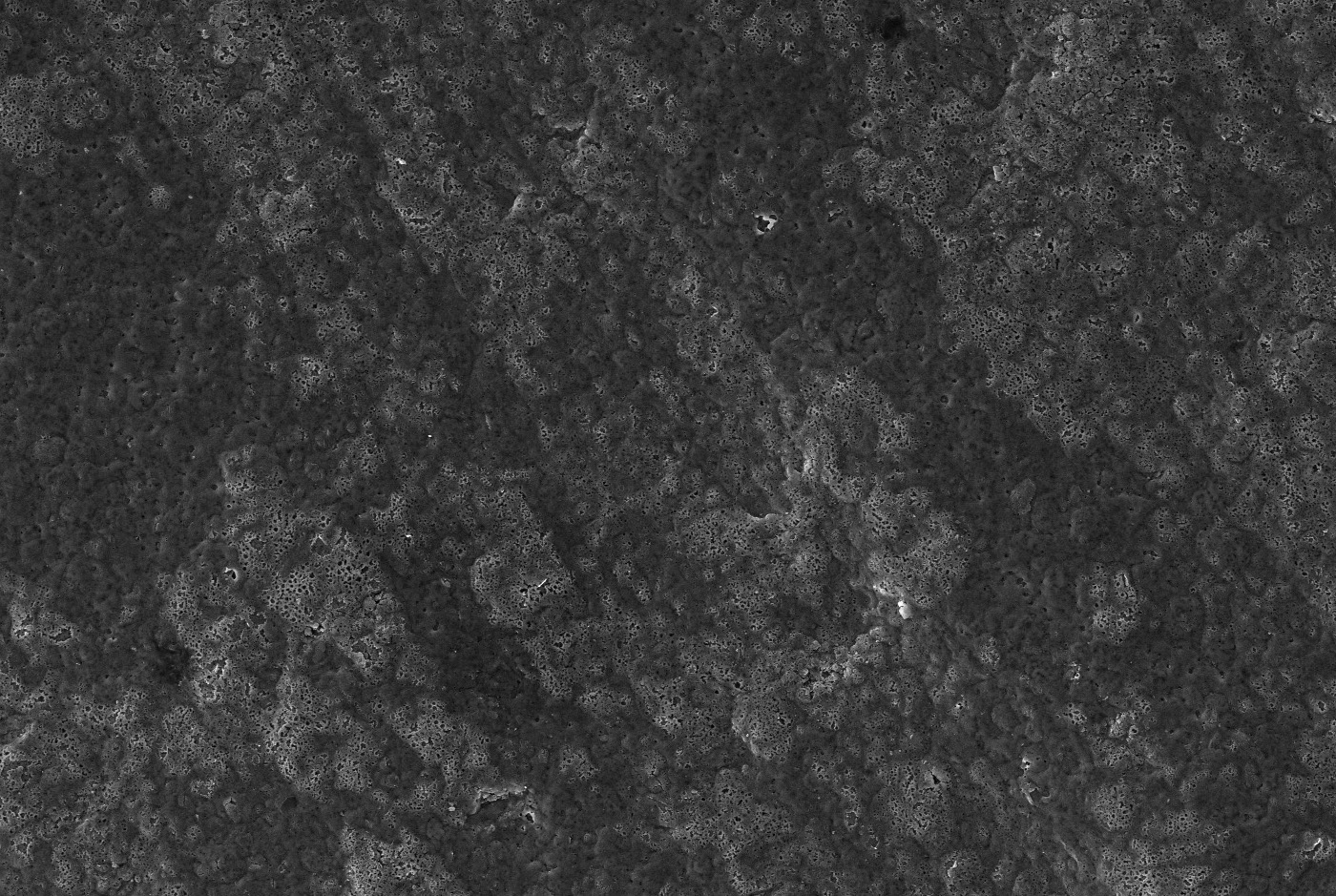}(8)\\		
	\end{tabular}
\caption{Titanium oxide dataset. Each image number corresponds to the same number of the experiments column presented in Table \ref{tab:matriz_de_planejamento}.}
\label{fig:dataset}
\end{figure*}

The samples used in this work were galvanostatic anodic titanium oxide ones prepared in oxalyc acid solution. In this electrochemical preparation method, a titanium plate is the anode in a two electrode electrochemical cell. A platinum plate was used as cathode. Then, the anode is polarized under constant current condition and an oxide film starts to form over the anode following the equation:
 $Ti+2H_{2}O\rightarrow TiO_{2}+4H^{+}+4e^{-}$
% Mexi nesta linhas em volta deste ponto
It is important to stress that $TiO_{2}$ is formed by the direct reaction between the metal and water over the metal. The surface morphology of the oxide is sensible to the  experimental conditions used. In the present case,  it were used different values of applied current, oxalyc acid concentration and solution temperature as described in Table \ref{tab:matriz_de_planejamento}.  Using a $2^3$ factorial design, it were performed 8 titanium oxide anodizations, generating, therefore, 8 classes of samples. From each class, it were acquired images from 8 different regions on their surface. Therefore, we have a total of 64 samples to be used in the model building. Each sample is a rectangular piece of the plate, which is measured through SEM-FEG technique generating a matrix (image) with a resolution of 3072$\times$2060 pixels. Figure \ref{fig:dataset} shows one image per class, illustrating the general aspect of the dataset.

\begin{table*}[!htpb]
\caption{Experimental Matrix  for the anodic titanium oxide samples preparation using a factorial design procedure}
\label{tab:matriz_de_planejamento}
\centering
\small
\begin{tabular}{cccc}
\hline 
Experiments & current density / $mA\: cm^{-2}$ & Temperature / $^{\circ}$C & Concentration / $mol\: L{}^{-1}$\\
\hline 
\hline 
1 & 10 & 10 & 0.05\\
\hline 
2 & 20 & 10 & 0.05\\
\hline 
3 & 10 & 30 & 0.05\\
\hline 
4 & 20 & 30 & 0.05\\
\hline 
5 & 10 & 10 & 0.5\\
\hline 
6 & 20 & 10 & 0.5\\
\hline 
7 & 10 & 30 & 0.5\\
\hline 
8 & 20 & 30 & 0.5\\
\hline 
\end{tabular}
\end{table*}

\section{Results}
\subsection{Fractal Theory}

Fractals are objects formally defined as a set of points whose Hausdorff-Besicovitch dimension (see the concept below) exceeds strictly the Euclidean dimension. In practice, it is an object generated through a dynamic system that presents infinite complexity and self-similarity \citep{M68}. Here, complexity states for the level of details under different scales. Self-similarity expresses the fact that if one takes the fractal under different observation scales, he will observe the repetition of patterns only changed by simple geometrical transforms (affine transforms). It is important to stress out that these properties are observed at infinite scale. In nature, we can find a lot of objects which present self-similarity and complexity at certain levels interval. Then, this is a strong motivation to approximate these structures through fractal metric. The most used of such fractal metrics is the one called fractal dimension.

\subsubsection{Fractal Dimension}

The first concept presented in the literature of fractal dimension \citep{M68} is the same one as Hausdorff dimension, which is based on the Hausdorff measure $\mathfrak{H}_{\delta}^{s}$ concept. Let $A$ be a subset of $\Re^n$ and $s$ and $\delta$ non-negative real values. Then:
\[
	\mathfrak{H}_{\delta}^{s}(A) = \inf\{\sum{\|U_i\|^s \mbox{such that $\{U_i\}$ is a $\delta$-cover of $A$}}\},
\]
where $\{U_i\}$ is a $\delta$-cover of $A$ if $A \subset \bigcup_{i=1}^{\infty}U_i$, being $0 < \|U_i\| \leq \delta$.

The Hausdorff s-measure $\mathfrak{H}^s$ is given by:
\[
	\mathfrak{H}^s(A) = \lim_{\delta \rightarrow 0}\mathfrak{H}_{\delta}^{s}(A).
\]

An interesting and important characteristic of this measure is that $\mathfrak{H}^{s}(A)$ is always $0$ for any $s < d_H$ and $\infty$ for any $s > d_H$. The real value $d_H$ is the so-called Hausdorff dimension of $A$, that is:
\[
	d_{H}(X) = \inf \left\{ s|H^{s}(X)=0 \right\} = \sup \left\{ s|H^{s}(X)=\infty \right\}.
\]

A special case of fractal dimension definition is the Bouligand-Minkowski (BM) dimension, decribed in the following section.

\subsubsection{Bouligand-Minkowski}

As well as in the Hausdorff dimension, BM also has an associated measure, which is, in this case, an upper measure $\overline{q_{\tau}}$ and a lower one $\underline{q_{\tau}}$ defined through:
\[
	\overline{q_{\tau}}(X,R) = \liminf_{r \rightarrow 0}q_{\tau}(X,R,r),
\]
\[
	\underline{q_{\tau}}(X,R) = \limsup_{r \rightarrow 0}q_{\tau}(X,R,r),
\]
where
\[
	q_{\tau}(X,R,r) = \frac{V(\partial X \oplus rR)}{r^{n-\tau}},
\]
with $-\infty < \tau < \infty$, $r < 0$, $\partial X$ is the boundary of $X$ and $\oplus$ denotes the morphological dilation by an element $R$ with radius $r$.

The upper and lower dimension, $\overline{D_B}$ and $\underline{D_B}$ respectively, are defined by:
\[
	\overline{D_B}(X,R) = \inf\{\tau | \overline{q_{\tau}}(X,R) = 0\},
\]
\[
	\underline{D_B}(X,R) = \inf\{\tau | \underline{q_{\tau}}(X,R) = 0\}.
\]

In a discrete space, like that of digital images here analyzed, the direct application of above equations is not viable. In such situations, a common practice is to employ neighborhood techniques. Particularly, here, we are interested in the estimation of dimension in $\Re^3$. In this space, the BM dimension may be presented through:
\[
	D_B(X) = \lim_{r \rightarrow 0}(3 - \frac{\log(V(\partial X \oplus Y_r))}{\log r}),
\]
where $V$ is the volume of the dilated structure and $Y_r$ is an Euclidean sphere with radius $r$.

\subsubsection{Fractal Descriptors}

As described in the introduction section, the fractal dimension is a scalar value and it is not enough to characterize such complex structures as those presented in Figure \ref{fig:dataset}. Therefore, we have developed an improved concept related to fractal dimension which is not a real number but a vector \citep{FB11}. This mathematical object was called fractal descriptor and its fundamental aspects is described in the following paragraphs \citep{BPFC08,FCB10,K99,PPFVOB05}.

We propose to extract the morphological properties from the samples through an analysis based on fractal geometry of the FEG images. In this way, an initial procedure is to estimate the fractal dimension of the data represented in FEG samples using the above Bouligand-Minkowski method due to its precision \citep{T95,PPFVOB05,BCB09}. We applied a neighborhood approach once the FEG image is described in a discrete space.

\begin{figure}[!htb] % Figuras lado a lado
	 \centering
	\includegraphics[width=0.29\columnwidth]{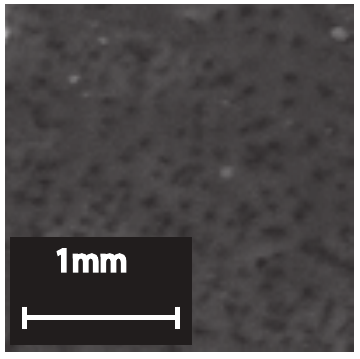}
	\includegraphics[width=0.69\columnwidth]{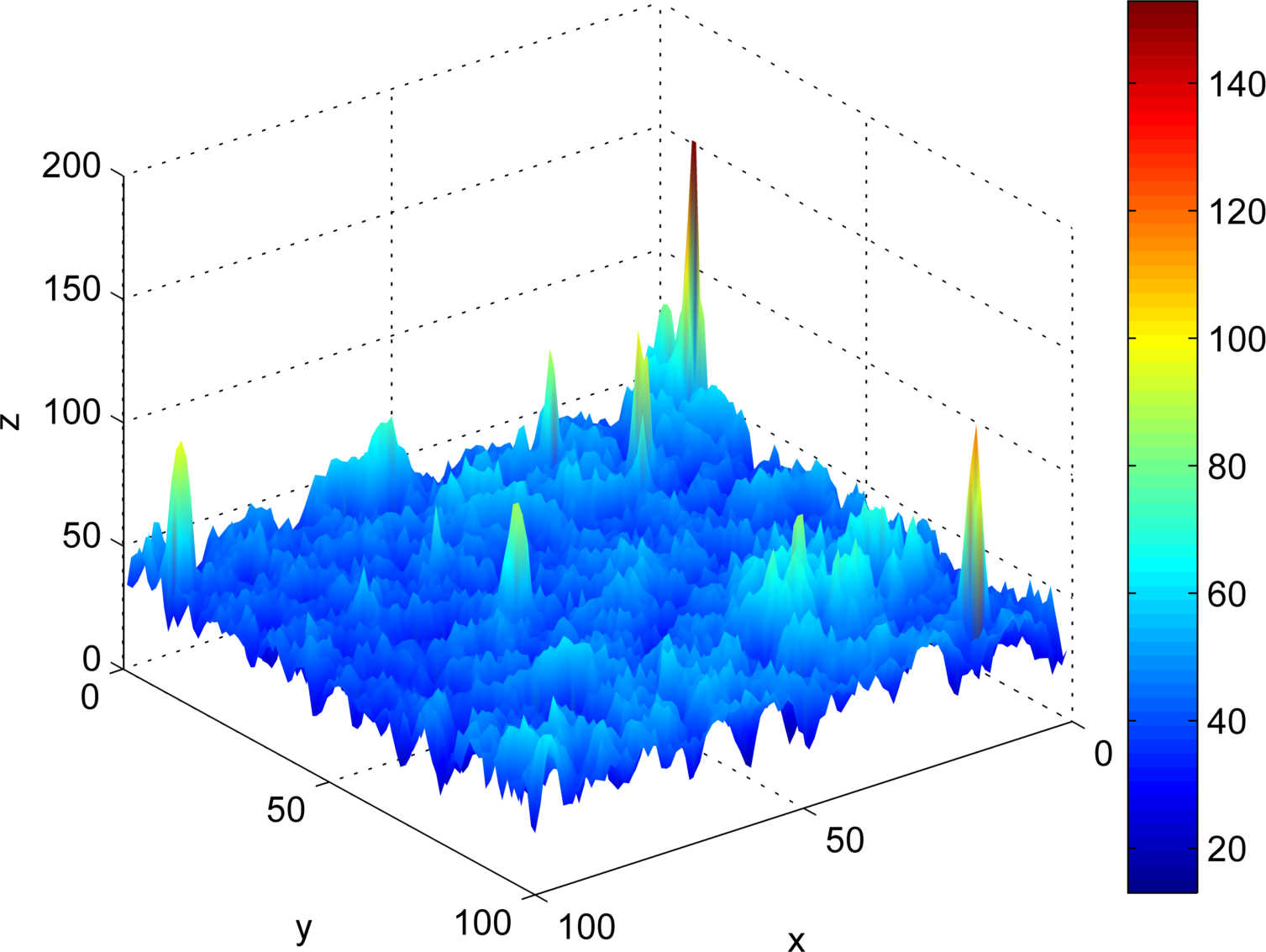}
          \caption{Texture image mapped onto a 3D surface. (a) Plain image. (b) Surface, obtained considering the gray levels as z axis.}
           \label{fig:BM}                                  
\end{figure}

The most intuitive way of calculating fractal dimension of a FEG image is to map the data onto a 3D gray intensity surface. The Figure \ref{fig:BM} shows an example of the image presented as a 3D gray  surface. This is performed in a simple manner by representing the image $I \in [1:M] \times [1:N] \rightarrow \Re$ in the surface $S$
\begin{equation}
	S = \{i,j,f(i,j)|(i,j) \in [1:M] \times [1:N]\},
\end{equation}
and 
\begin{equation}
	f(i,j) = \{1,2,...,max\_{value}\}|f = I(i,j),
\end{equation}
being $max\_{value}$ the maximum value in the FEG data.

\begin{figure}[!htb] % Figuras lado a lado
	 \centering
	 \includegraphics[width=0.3\textwidth]{surf_v3.png}
	 \includegraphics[width=0.3\textwidth]{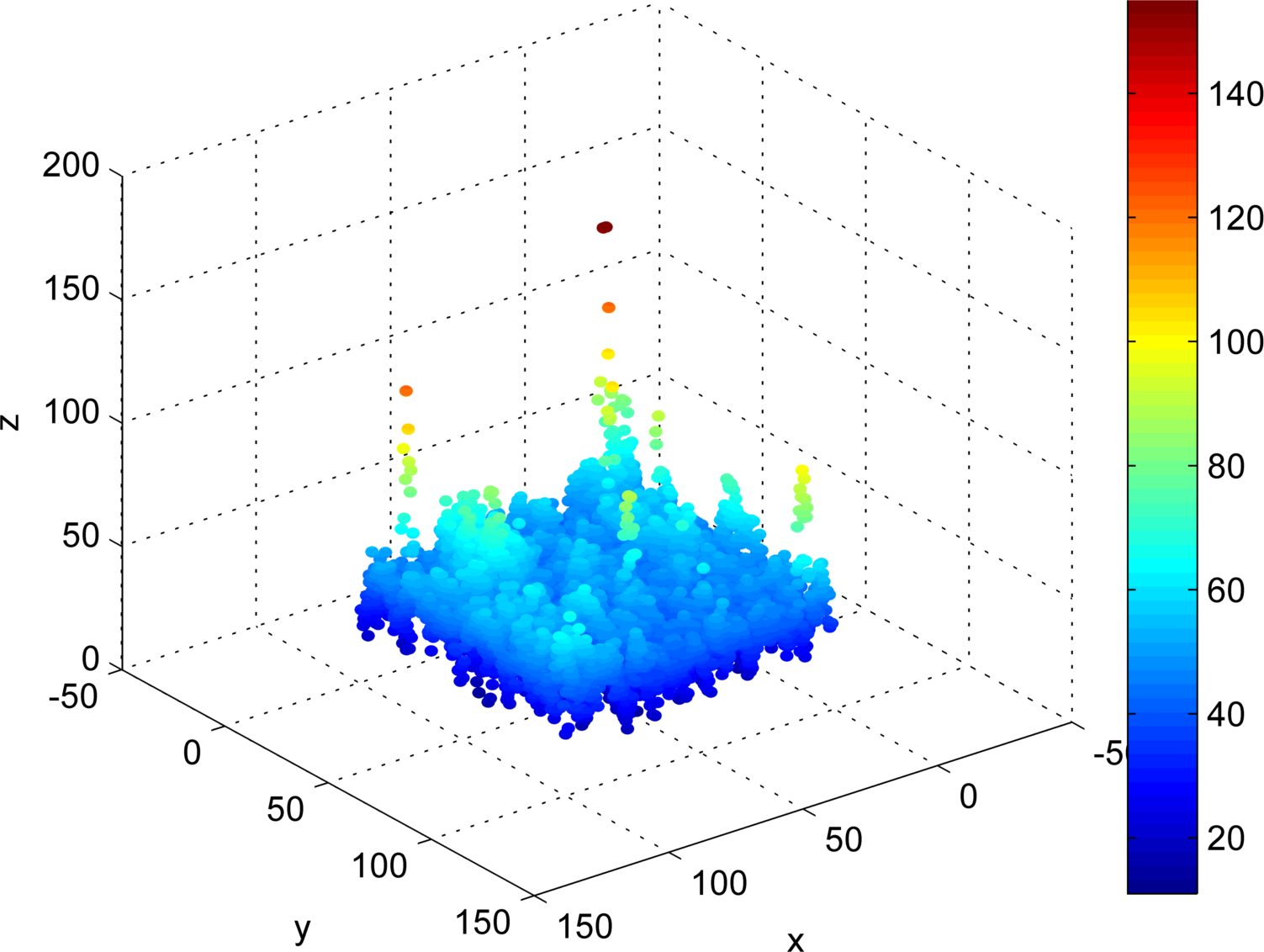}
	 \includegraphics[width=0.3\textwidth]{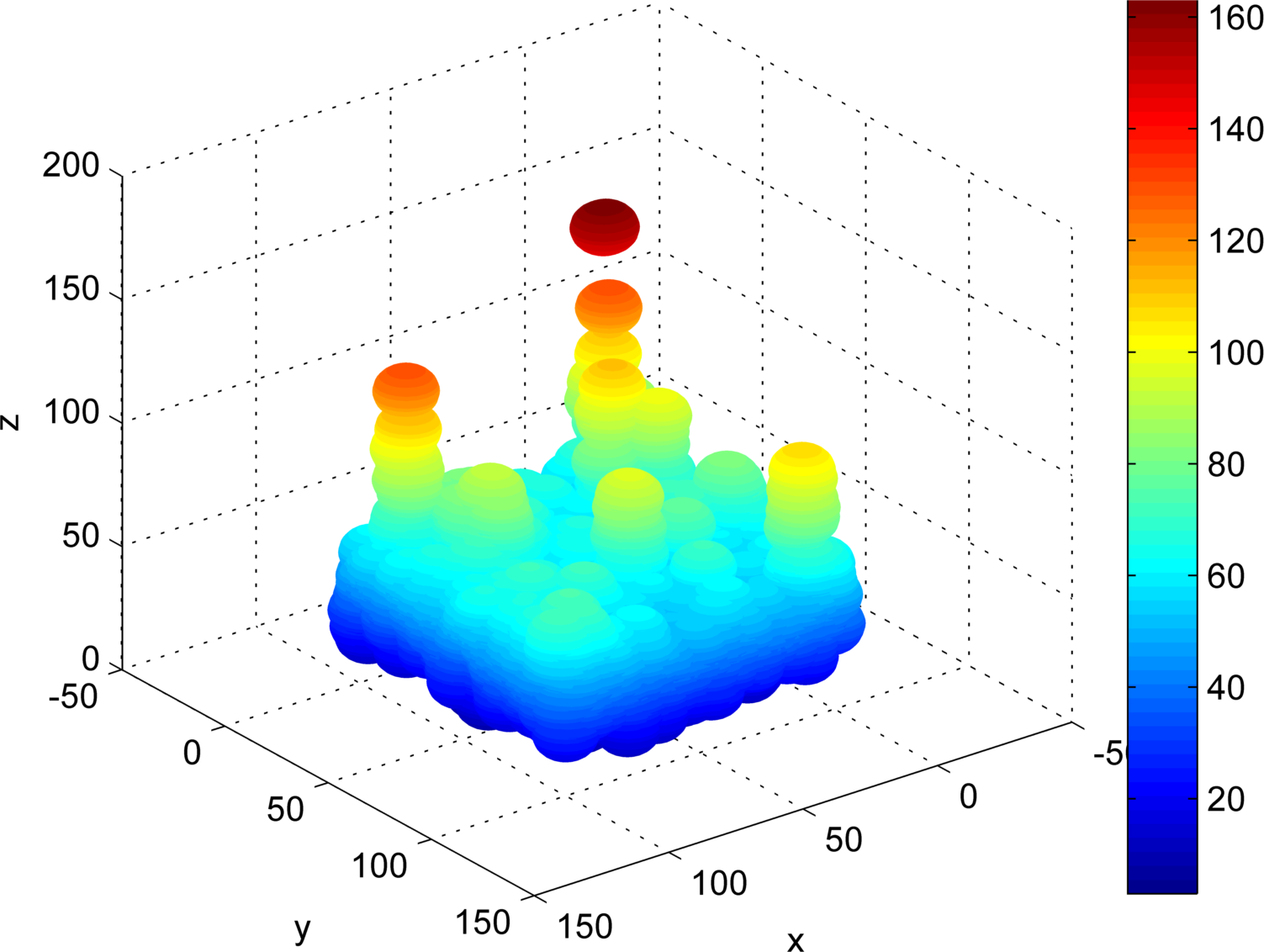}
           \caption{Dilated surfaces with different radii. (a) Original surface. (b) Radius 2. (c) Radius 10.}
           \label{fig:dilat}                                  
   \end{figure}

A classical way of calculating the BM dimension of a surface in discrete space is dilating it with spheres varying the radius $r$, which is presented in Figure \ref{fig:dilat}. This Figure illustrates the dilation process for two different values of $r$. Using this approach, we compute the dilation volume $V(r)$ (number of points inside the dilated surface) for each value of $r$. The dimension is given through:
\[
	D_B = 3 - \lim_{r \rightarrow 0}\frac{\log(V(r))}{\log(r)}
\]
For computational efficiency purposes, an usual solution to calculate $V(r)$ is to employ the exact Euclidean Distance Transform (EDT) \citep{FCTB08}, which is for a surface $S$ in three dimensions given by:
\begin{equation}
	EDT(p) = \min\{d(p,q)|q \in U \backslash S\},
\end{equation}
where $d$ is the Euclidean distance and $U \backslash S$ corresponds to the complement of $S$ taken over a cube $U$ (universal set) which contains $S$. When we are dealing with exact EDT, the distance has predefined values:
\begin{equation}
	E = {0,1,\sqrt{2},...,l,...},
\end{equation}
where
\begin{equation}
	l \in D = \{d | d = (i^2+j^2)^{1/2};i,j \in \mathbb{N}\}
\end{equation}

Initially, we define the set $g_r(S)$ of points at a distance $r$ from $S$:
\begin{equation}
	 g_r(S) = \left\{(x,y,z)|[(x-S_x)^2 + (y-S_y)^2 \\
	 + (z-S_z)^2]^{1/2} = E(r) \right\},
\end{equation}
where $S_x$, $S_y$, $S_z$ are the coordinates of points in $S$.

The dilation volume $V(r)$ is given by:
\begin{equation}
	V(r) = \sum_{i=1}^{r}{Q(i)},
\end{equation}
where $Q(i)$ is provided through the following expression:
\[
	Q(i) = \sum_{(x,y,z) \in U}{\chi_{g_r}(x,y,z)},
\]
where $\chi$ states for the indicator function.

Therefore, $V(r)$ and $r$ present a power law relation. The descriptors themselves are obtained using the following relation $u$:
\[
	u:\log(r) \rightarrow \log(V(r)),
\]
where $V$ acts as a fractality measure and $r$ is the scale parameter. 

\section{Discussion}

Several times, to extract the information from function $u$, it is necessary a mathematical space transformation, such as Fourier or Principal Component transform. 

\begin{figure*}[!htpb]
\centering
\includegraphics[width=0.8\textwidth]{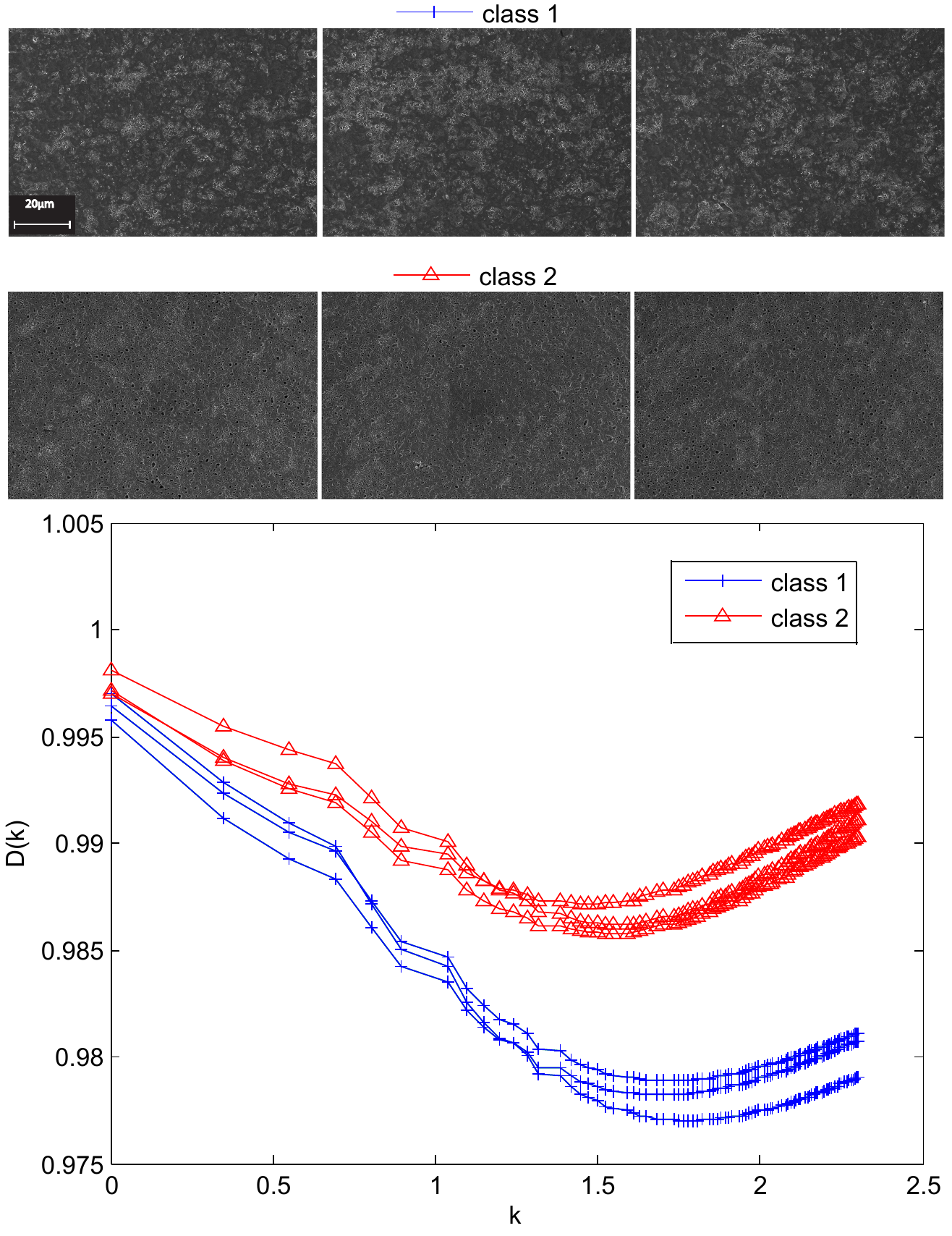}
\caption{Discrimination performance of proposed descriptors.}
\label{fig:classes}
\end{figure*}

Here, we apply the Principal Component  (PC) transform \citep{DH00}  of the fractal descriptors in the  classification  of the different investigated materials. Figure \ref{fig:classes} illustrates the ability of discrimination through BM fractal descriptors.

The performance of the proposed approach is verified in a task of classification of titanium oxide films prepared using different experimental conditions as described in the Section \ref{sec:mat}. Each condition corresponds to one class.

The fractal descriptors are extracted from the images and the Principal Component Analysis (PCA) is applied over each set of fractal descriptors. The components are thus classified by Linear Discriminant Analysis (LDA) \citep{DH00}, using leave-one-out cross-validation procedure. To check the quality of the results, they were compared to two classical texture analysis methods, that is, Fourier \citep{GW02} and histogram entropy \citep{GW02}. To guarantee that the comparison is valid, the same PCA-LDA methodology was applied also with these last methods.

\begin{figure}[!htpb]
\centering
\includegraphics[width=\columnwidth]{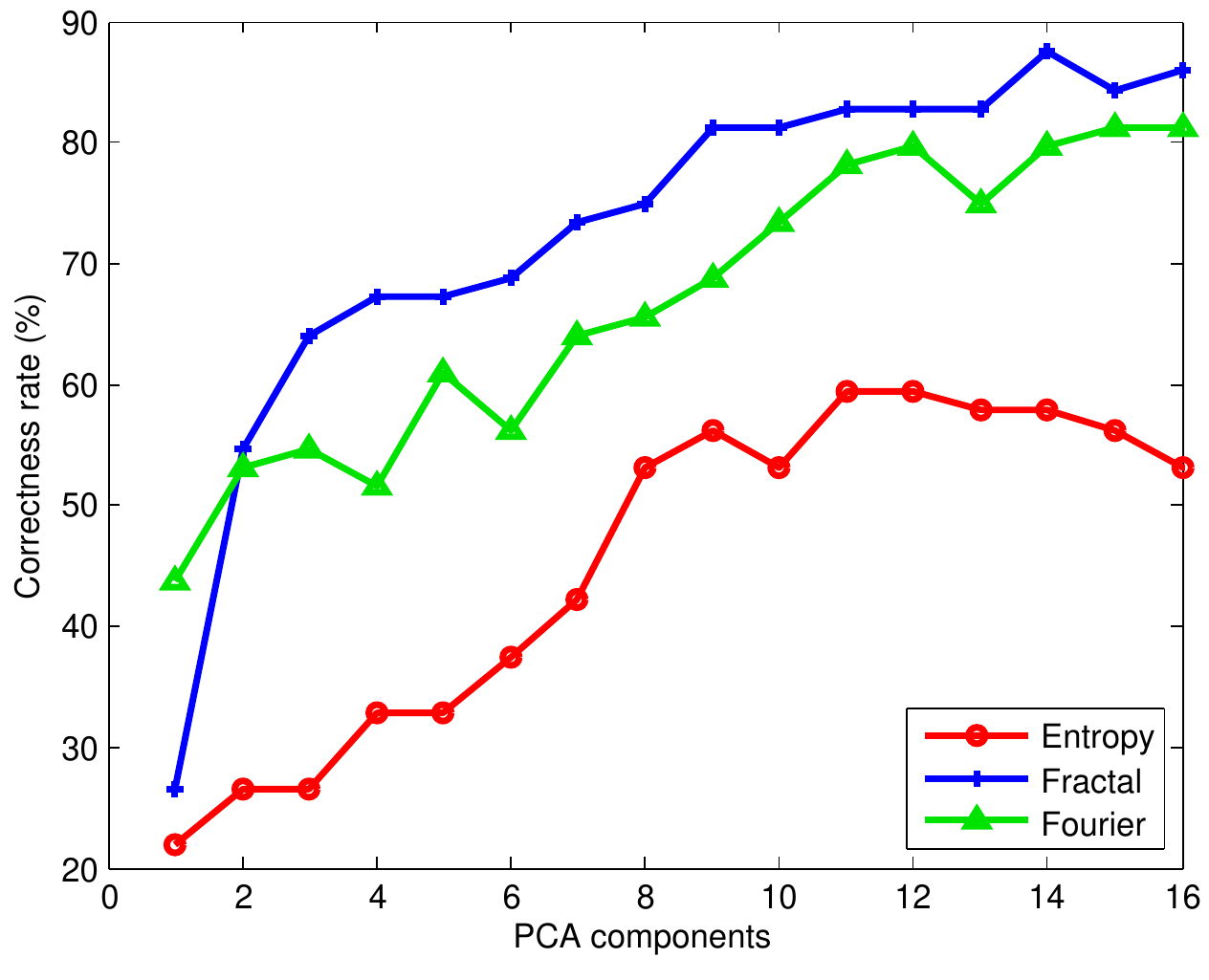}
\caption{Correctness rate according to the number of PC components in each compared method.}
\label{fig:pca}
\end{figure}

Figure \ref{fig:pca} illustrates the behavior of PC components relative to their correctness rate in the classification process for each methods used. It shows the number of components up to 16 to detect the necessary number of principal components to be used.

We observe that the rate of correctness increases as the number of components increases and the Fractal descriptors achieve its maximum values faster than the other methods used. In this case, it is necessary  to use 14 components. The results presented in the PC graph are summarized in the Table \ref{tab:result} showing the best result obtained by each descriptor, with the measure error and the number of components used to achieve such result. After  the maximum correctness, the performance decrease for every method which means that the use of more components in the LDA procedure only add noise to the set of features, damaging the performance of the descriptors. 

\begin{table}[!htpb]
	\centering
	\small
		\begin{tabular}{ccc}
			\hline
                 Method & Correctness Rate (\%) & Number of descriptors\\
                 \hline
                 \hline
Entropy & 59.38 $\pm$ 0.01 & 11\\
Fourier & 81.25 $\pm$ 0.01 & 15\\
Fractal & 87.50 $\pm$ 0.01 & 14\\
								 \hline			
		\end{tabular}
	\caption{Correctness rate for each compared descriptor.}
	\label{tab:result}
\end{table}

Now, we see that fractal descriptors showed the best correctness rate result with a 8\% of advantage over the second best approach, Fourier descriptors. This result was expected due to the intrinsic ability of fractal descriptors in extracting a rich information of the topography of the surface. Such topography is directly related to important physical aspects of the material, like roughness, grain boundaries, morphologic defects, reactivity and total surface area. In this sense, we observe that the proposed descriptors are capable of accurately capturing nuances which are essential in the discrimination skill of our vision system. It is important to notice, however, that the use of descriptors associated to the classifier computational method turns possible the analysis of details with a complex processing framework which ensures a more precise and robust solution for the discrimination problem.

\section{Conclusions}

This work showed an application of fractal descriptors to the classification of samples of titanium oxide material under different experimental conditions.

We compared the performance of fractal to other classical texture descriptors approaches in the literature, to know, histogram entropy and Fourier. The results showed that fractal approach obtained the best result, providing the most accurate classification. We verified, as expected, that fractal geometry is a powerful tool to describe such nanoscale image. This is explained by the flexibility of fractals in modeling topographies arising from natural systems, like that present in titanium oxide samples.

The result encourages the research and enhancing of novel fractal-based approaches, applied to many challenging problems related to the discrimination and description of materials under different experimental conditions.

It is important to stress out that this procedure can be used also for different natural or synthetic samples which can be found in the literature from different areas of the knowledge.

\section{Acknowledgements}
\label{sec:Acknowledgements}
Jo\~ao B. Florindo is grateful to CNPq(National Council for Scientific and Technological Development, Brazil) for his doctorate grant. Ernesto C. Pereira and Mariana S. Sikora would like to thank the Brazilian research funding agencies CNPq and FAPESP (proc:2008/00180-0) for their financial support.
Odemir M. Bruno gratefully acknowledges the financial support of CNPq (National Council for Scientific and Technological Development, Brazil) (Grant \#308449/2010-0 and \#473893/2010-0) and FAPESP (The State of S\~ao Paulo Research Foundation) (Grant \# 2011/01523-1). 

%\bibliography{Oxalico_c}

\end{document}